\documentclass[preprint,aps,prb,showpacs,superscriptaddress]{revtex4}
\usepackage{multirow}
\usepackage{graphicx}
\usepackage{dcolumn}
\usepackage{bm}

\begin{document}

\title{
Structural and electronic properties of Sc$_n$O$_m$
($n$=1-3,$m$=1-2$n$) clusters: A theoretical study using screened
hybrid density functional theory}

\author{Yu Yang}
\affiliation{LCP, Institute of Applied Physics and Computational
Mathematics, P.O. Box 8009, Beijing 100088, People's Republic of
China}%
\author{Haitao Liu}
\affiliation{LCP, Institute of Applied Physics and Computational
Mathematics, P.O. Box 8009, Beijing 100088, People's Republic of
China}%
\author{Ping Zhang}
\thanks{To whom correspondence should be
addressed. E-mail: zhang\_ping@iapcm.ac.cn (P.Z.)}%
\affiliation{LCP, Institute of Applied Physics and Computational
Mathematics, P.O. Box 8009, Beijing 100088, People's Republic of
China}%


\begin{abstract}

The structural and electronic properties of small scandium oxide
clusters Sc$_n$O$_m$ ($n$=1-3, $m$=1-2$n$) are systematically
studied within the screened hybrid density functional theory. It is
found that the ground states of these scandium oxide clusters can be
obtained by the sequential oxidation of small ``core'' scandium
clusters. The fragmentation analysis demonstrates that the ScO,
Sc$_2$O$_2$, Sc$_2$O$_3$, Sc$_3$O$_3$, and Sc$_3$O$_4$ clusters are
especially stable. Strong hybridizations between O-2$p$ and Sc-3$d$
orbitals are found to be the most significant character around the
Fermi level. In comparison with standard density functional theory
calculations, we find that the screened hybrid density functional
theory can correct the wrong symmetries, and yield more precise
description for the localized 3$d$ electronic states of scandium.

\end{abstract}

\maketitle

\section{Introduction}

Transition metal oxide clusters have attracted enormous attention
because their structural, electronic, and magnetic properties are
often quite different from those in their bulk phase
\cite{Nayak98,Reddy99,Harrwason00,Tono03,Pykavy04,Qu06,Uzunova08,WangQ08,Mowbray09,WangYB10}.
For example, small clusters like (ZnO)$_n$, (V$_2$O$_5$)$_n$,
(CrO$_3$)$_n$ form planar structures with very small sizes, and some
small clusters present novel magnetic properties
\cite{WangQ08,WangYB10,Vyboishchikov01,Li06}. For scandium (Sc),
continuous interests are shown for several reasons. Firstly, Sc is
the first transition metal in the periodic table, and thus is always
taken as a prototype to study the complex phenomena associated with
$d$ shell electrons \cite{Gonzales00}. Secondly, Sc oxide
nanostructures can be used as catalysts in some reactions like the
selective reduction of nitric oxides with methane
\cite{Gonzales00,Fokema98}. Thirdly, some Sc oxide clusters have
been recognized in the spectra of M-type stars
\cite{Gonzales00,Merrill62}. Finally, some Sc oxide clusters can be
steadily encapsulated into closed carbon cages of fullerenes to form
a novel nanostructure \cite{Stevenson08,Valencia09,Chaur09}.

The small Scandium oxide clusters can be prepared by laser ablation
of scandium metal in the presence of oxygen-saturated atmosphere
\cite{Chertihin97,Wu98,Kushto99,Zhao11}. With the presence of oxygen
or nitrogen oxide gases, scandium oxide clusters ranging from ScO to
Sc$_3$O$_6$ have already been generated and detected \cite{Zhao11}.
Many experimental studies have already been applied to study the
structural, energetic, vibrational, electronic, and magnetic
properties of scandium oxide clusters. For example, the
photoionization spectra of Sc$_n$O were measured and strongly
size-dependent ionization potentials were observed
\cite{Zhao11,Gutsev00}. For the ScO molecule, the electron-spin
resonance and optical spectroscopy in neon and argon matrices
revealed that its ground state is a doublet \cite{Zhao11,Weltner67},
and the molecular bond length and vibrational frequency have been
measured to be 1.668 \AA~ and 965 $cm^{-1}$ respectively
\cite{WangYB10,Huber79}. Accompanying with these experimental
results, lots of theoretical calculations, especially those based on
density functional theory (DFT), were also carried out to explore
the ground-state electronic structures of scandium oxide clusters.
However, most of those studies on Sc oxide clusters were mostly on
the standard DFT level, within the local density approximation (LDA)
or generalized gradient approximation (GGA). In this paper, we
systematically study the atomic and electronic structures of Sc
oxide clusters using the screened hybrid density functional theory,
which has proven to be able to significantly improve the description
of finite, molecular systems, such as atomization energies, bond
lengths and vibrational frequencies using standard DFT theories
\cite{Nieminen09}.

\section{Computational Method}

For extended systems, the ground-state electronic properties can be
obtained by solving the Kohn-Sham equation within DFT
\cite{Hohenberg64,Kohn65}, utilizing the standard approximations for
the exchange-correlation ($xc$) energy functional $E_{xc}$, i.e.,
LDA and GGA. One problem of the two most commonly applied
functionals is that they rely on the $xc$ energy per particle of the
uniform electron gas, and thus are expected to be useful only for
systems with slowly varying electron densities (local/semilocal
approximation) \cite{Hummer09}. Although the LDA and GGA functionals
have proven to be rather universally applicable in theoretical
materials science and achieve fairly good accuracy for ionization
energies of atoms, dissociation energies of molecules, and cohesive
energies, as well as bond lengths and geometries of molecules and
solids \cite{Hummer09,Gunnarsson76,Jones89,Kohn99}, this
unexpectedly good performance for the ground-state properties of
many materials is believed to be due to the partial error
cancelation in the exchange and correlation energies
\cite{Hummer09}. In order to improve the performance of DFT on a
theoretical basis, one approach is to add a portion of the non-local
Hartree-Fock (HF) exchange to the local/semilocal approximation for
exchange-correlation functional. In computational solid state
physics, a breakthrough was achieved by Heyd, Scuseria, and
Ernzerhof, who proposed the HSE03 functional defined by
\cite{Heyd03,Heyd04,Heyd06}:
\begin{equation}
    E_{xc}^{\rm HSE03}=\frac{1}{4}E_{x}^{\rm HF}({\rm SR})+\frac{3}{4}E_{x}^{\rm PBE}({\rm SR})+E_{x}^{\rm PBE}({\rm LR})+E_{c}^{\rm PBE}.
\end{equation}
The HSE03 functional benefits from the range separation of the HF
exchange into a short-(SR) and long-range (LR) contribution and
replaces the latter by the corresponding DFT exchange part. The
HSE03 functional as well as its corrected form the HSE06 functional
\cite{Krukau06}, have been extensively applied to calculate
ground-state properties of solid and molecular systems, as well as
adsorption energies for molecules, and have proven to be better than
the LDA and GGA functionals in describing finite molecular systems.

The results presented in this work are obtained using the projector
augmented wave (PAW) method \cite{PAW} as implemented in the Vienna
ab initio simulation package (VASP) \cite{VASP}. For standard DFT
calculations the $xc$ energy is treated within the GGA using the
parameterization of Perdew, Burke, and Ernzerhof (PBE)
\cite{PBE1,PBE2} to compare with previously published calculations.
In the present HSE calculations, the HSE06 hybrid functional
\cite{Krukau06} is applied, where the range separation parameter is
set to 0.2 \AA$^{-1}$ for both the semilocal as well as the nonlocal
part of the exchange functional. From now on, this particular
functional will be referred to as HSE. A plane wave basis set with a
cutoff energy of 400 eV is adopted, and the scandium 3{\it
s}$^2$3{\it p}$^6$4{\it s}$^2$3{\it d}$^1$ and oxygen 2{\it
s}$^2$2{\it p}$^4$ electrons are treated as fully relaxed valence
electrons. A Fermi broadening \cite{Weinert1992} of 0.05 eV is
chosen to smear the occupation of the bands around the Fermi energy
(E$_f$) by a finite-$T$ Fermi function and extrapolating to $T=0$ K.
The supercell containing the scandium oxide clusters is chosen to be
without symmetry, and the cell size along each direction is larger
than 15 \AA. A quasi-Newton algorithm is used to relax the scandium
and oxygen ions for all scandium oxide clusters, with the force
convergence criteria of 0.01 eV/\AA~ in PBE calculations and 0.03
eV/\AA~ in HSE calculations.

\section{Results and Discussions}

\subsection{Geometrical structures}

Firstly, the ScO and two isomers of ScO$_2$ are shown in Fig.1 (a)
and (b) respectively. The optimized bond length of ScO is 1.677 \AA~
in PBE calculation, and 1.659 \AA~ in HSE calculation. The PBE
result is in good agreement with former GGA calculations
\cite{WangYB10}. The HSE06 result is near the ones obtained from the
SCF \cite{Dolg87} and hybrid B1LYP calculations
\cite{Qu06,Uzunova08}. Both GGA and HSE06 calculations agree well
with the experimental bond length (1.668\AA) \cite{Huber79}.
Moreover, The vibrational frequency of ScO is 1056 $cm^{-1}$ in HSE
calculation, which is near the experimental value of 965 $cm^{-1}$.

\begin{figure}[tbp]
\begin{center}
\includegraphics[width=0.8\linewidth]{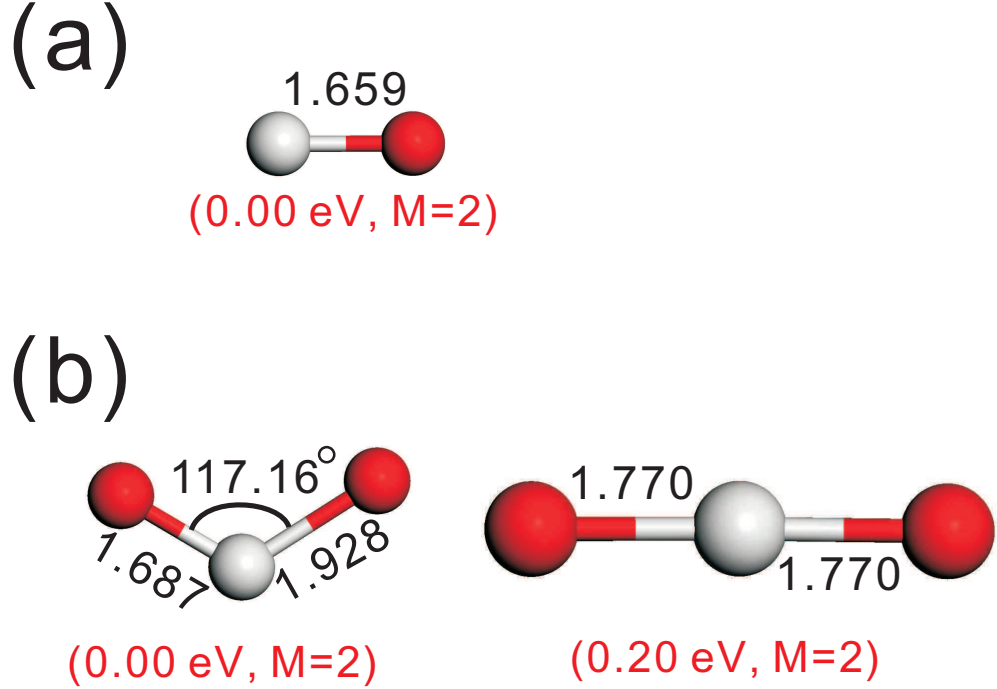}
\end{center}
\caption{(Color online). The low-energy structures of (a) ScO and
(b) ScO$_2$ clusters. Grey and red balls represent the scandium and
oxygen atoms, respectively. The bond lengths are in units of \AA.
The number in parenthesis is the relative energy (in eV) with
respect to the corresponding ground state. Note that the spin
multiplicity is also given in the parenthesis.}
\end{figure}%

For ScO$_2$, the obtained ground state is the obtuse triangle
structure, with the apex angle of 117.16$^{\circ}$. It is
interesting to note that these two Sc-O bonds are not equivalent and
the bond lengths are 1.687 and 1.928 \AA respectively. In PBE
calculations, the triangle is isosceles, with the apex angle of
126.19$^{\circ}$ and the ScO bond length of 1.781 \AA~, in agreement
with that have been found by using the DFT/PBE \cite{WangYB10} and
DFT/BPW91 methods \cite{Gutsev00}. It is clear that the HSE06
calculation reduces the symmetry of the ground state. Both GGA and
HSE06 demonstrates the doublet is the most stable electronic state
for the ground state of the ScO$_2$ cluster, in good agreement with
the experimental result. The next stable state of ScO$_2$ is the
O-Sc-O linear structure with 0.2~eV higher than the ground state.

\begin{figure}[tbp]
\begin{center}
\includegraphics[width=0.8\linewidth]{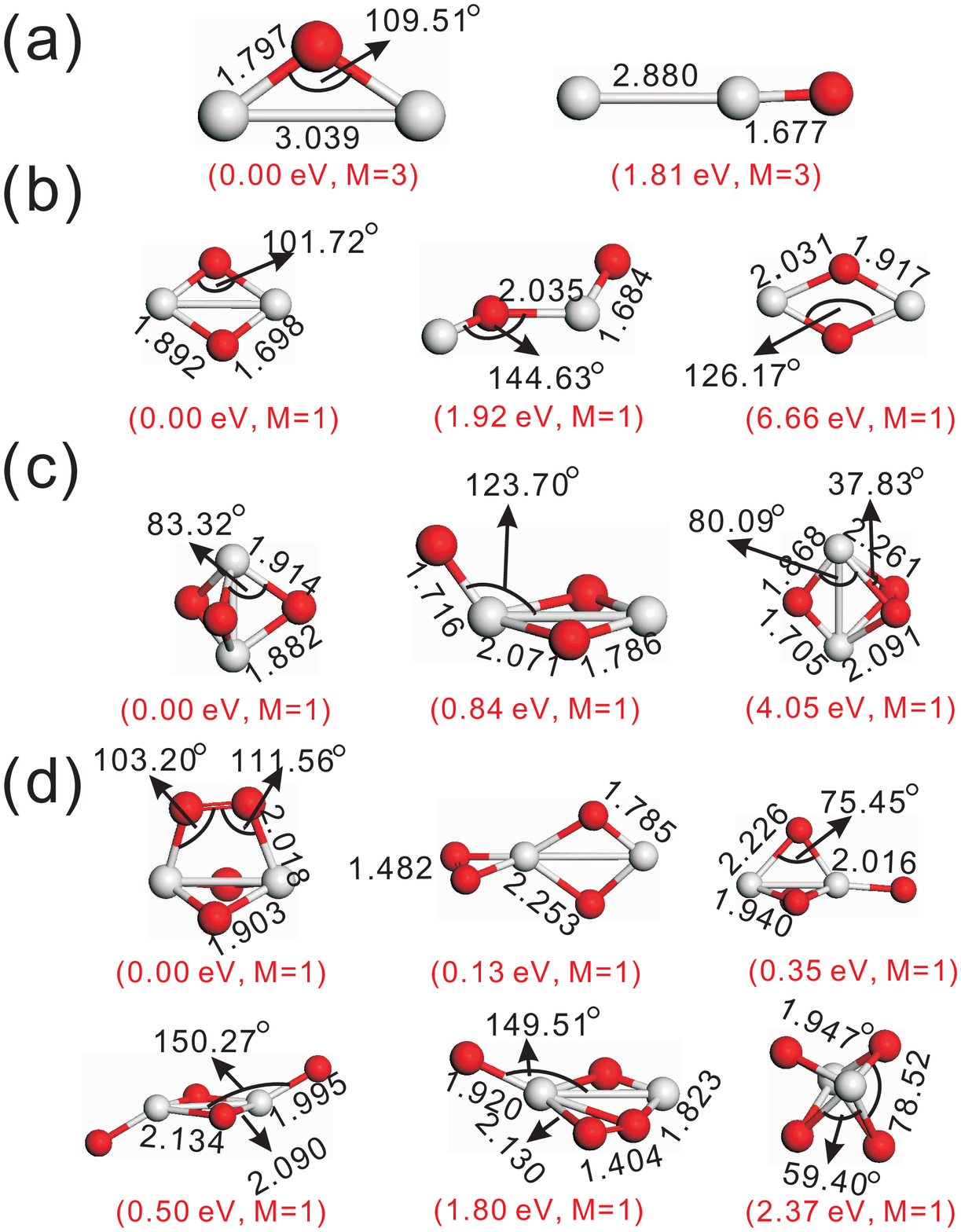}
\end{center}
\caption{(Color online). The low-energy structures of (a) Sc$_2$O,
(b) Sc$_2$O$_2$, (c) Sc$_2$O$_3$, and (d) Sc$_2$O$_4$ clusters. Grey
and red balls represent the scandium and oxygen atoms, respectively.
The bond lengths are in units of \AA. The number in parenthesis is
the relative energy (in eV) with respect to the corresponding ground
state. Note that the spin multiplicity is also given in the
parenthesis.}
\end{figure}%

The low-lying isomers of Sc$_2$O$_m$ (m=1$\sim$4) are presented in
Fig.2 (a)-(d). The ground-state structure of Sc$_2$O cluster is also
an obtuse triangle. The apex angle is 109.51$^{\circ}$, and the two
Sc-O bonds are both 1.797 \AA~ long, indicating that the cluster has
a $C_{2v}$ point symmetry. The ground state of Sc$_2$O is a triplet,
which is in agreement with previous DFT studies \cite{WangYB10}. The
next stable structure of Sc$_2$O is linear, and has a triplet
electronic state. The Sc$_2$O$_2$ cluster adopts a distorted square
as its ground-state structure. The Sc-O bond lengths are 1.892,
1.698, 1.892 and 1.698 \AA~ respectively. Since the PBE calculation
prefers the rhombus with D$_{2h}$ symmetry, our result is in
accordance with the rule that the HSE06 functional tends to reduce
the point group symmetry of some clusters obtained from the PBE
calculations. The ground-state of Sc$_2$O$_2$ is singlet, and the
next stable state is the triplet state of the same structure. The
other two structures of Sc$_2$O$_2$ in Fig. 2(b) actually have much
higher electronic free energies. For Sc$_2$O$_3$, the ground-state
structure is the singlet trigonal bi-pyramid structure shown in Fig.
2(c). This structure has a $D_{3h}$ symmetry in PBE calculation,
which disappears in HSE calculation. Until now, our PBE
calculational results are in good agreements with the results
reported by Wang {\it et al.} \cite{WangYB10}. The main modification
of the HSE method, as we can see from the above description, is
revealing a lower symmetry for the ground-state structure. In the
previous work by Wang {\it et al.}, the ground-state structure of
Sc$_2$O$_4$ is the one with a molecular O$_2$ adsorbing at one
corner of the rhombus Sc$_2$O$_2$ cluster, which is however, proven
to be the next stable structure for Sc$_2$O$_4$. The more stable
structure we revealed for Sc$_2$O$_4$ is the adsorbing structure
with an O$_2$ molecule adsorbing on top of the Sc-Sc bond, as shown
in Fig. 2(d). The ground and next stable states of Sc$_2$O$_4$ are
both singlet.

\begin{figure}[tbp]
\begin{center}
\includegraphics[width=0.8\linewidth]{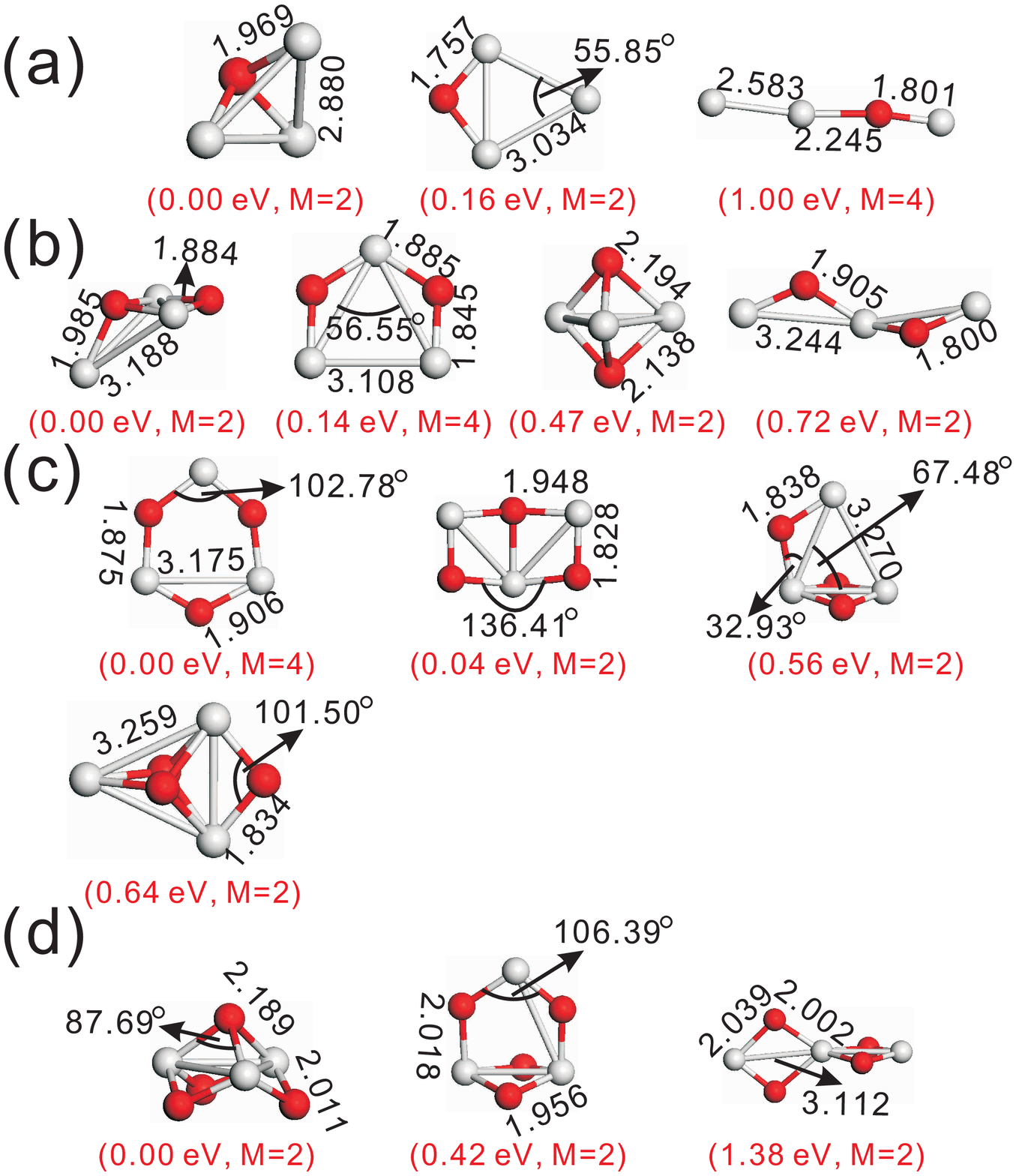}
\end{center}
\caption{(Color online). The low-energy structures of (a) Sc$_3$O,
(b) Sc$_3$O$_2$, (c) Sc$_3$O$_3$, and (d) Sc$_3$O$_4$ clusters. Grey
and red balls represent the scandium and oxygen atoms, respectively.
The bond lengths are in units of \AA. The number in parenthesis is
the relative energy (in eV) with respect to the corresponding ground
state. Note that the spin multiplicity is also given in the
parenthesis.}
\end{figure}%

The Sc$_3$O cluster has been reported to have a singlet ground
state, with the structure of an oxygen atom adsorbing on top of an
equilateral scandium triangle \cite{WangJL09}. Our result of Sc$_3$O
is the same with previous, in both PBE and HSE calculations. The
next stable structure of Sc$_3$O is the one with an oxygen atom
adsorbing at the bottom of an isosceles scandium triangle, having a
singlet electronic state and a free electronic energy of 0.16 eV
higher than that of the ground-state of Sc$_3$O. Except for Sc$_3$O,
other Sc$_3$O$_m$ clusters have seldom been studied before. Only a
systematic study on the Sc$_3$O$_6$ cluster is reported very
recently \cite{Zhao11}. The ground-state of Sc$_3$O$_2$ is the
doublet state of the adsorbing structure of a scandium atom on the
Sc$_2$O$_2$ cluster. We also notice that there are several structure
for the Sc$_3$O$_2$ cluster having close free electronic energies,
which are thus shown in Fig. 3(b). They are the pentagon, bi-pyramid
and bi-triangle structures respectively. The Sc$_3$O$_3$ cluster
adopts a hexagon structure with a fourfold spin multiplicity as its
ground state. The smallest O-Sc-O angle is 101.59$^{\circ}$ in PBE
calculation, and 102.78$^{\circ}$ in HSE calculation. We also listed
several other structures of Sc$_3$O$_3$ in Fig. 3(c), especially the
next stable doublet state of the folded rectangular structure, which
is only 0.04 eV higher in free electronic energy. For Sc$_3$O$_4$,
the ground-state structure can be seen as the adsorption of an
oxygen atom at the center of the hexagonal Sc$_3$O$_3$ cluster, or
the adsorption of three oxygen atoms at the three bottom sides of
the pyramid Sc$_3$O cluster. The next stable structure of
Sc$_3$O$_4$, which is 0.42 eV higher in free electronic energy, can
be seen as replacing one oxygen atom in the hexagonal Sc$_3$O$_3$
cluster with two oxygen atoms.

\begin{figure}[tbp]
\begin{center}
\includegraphics[width=0.8\linewidth]{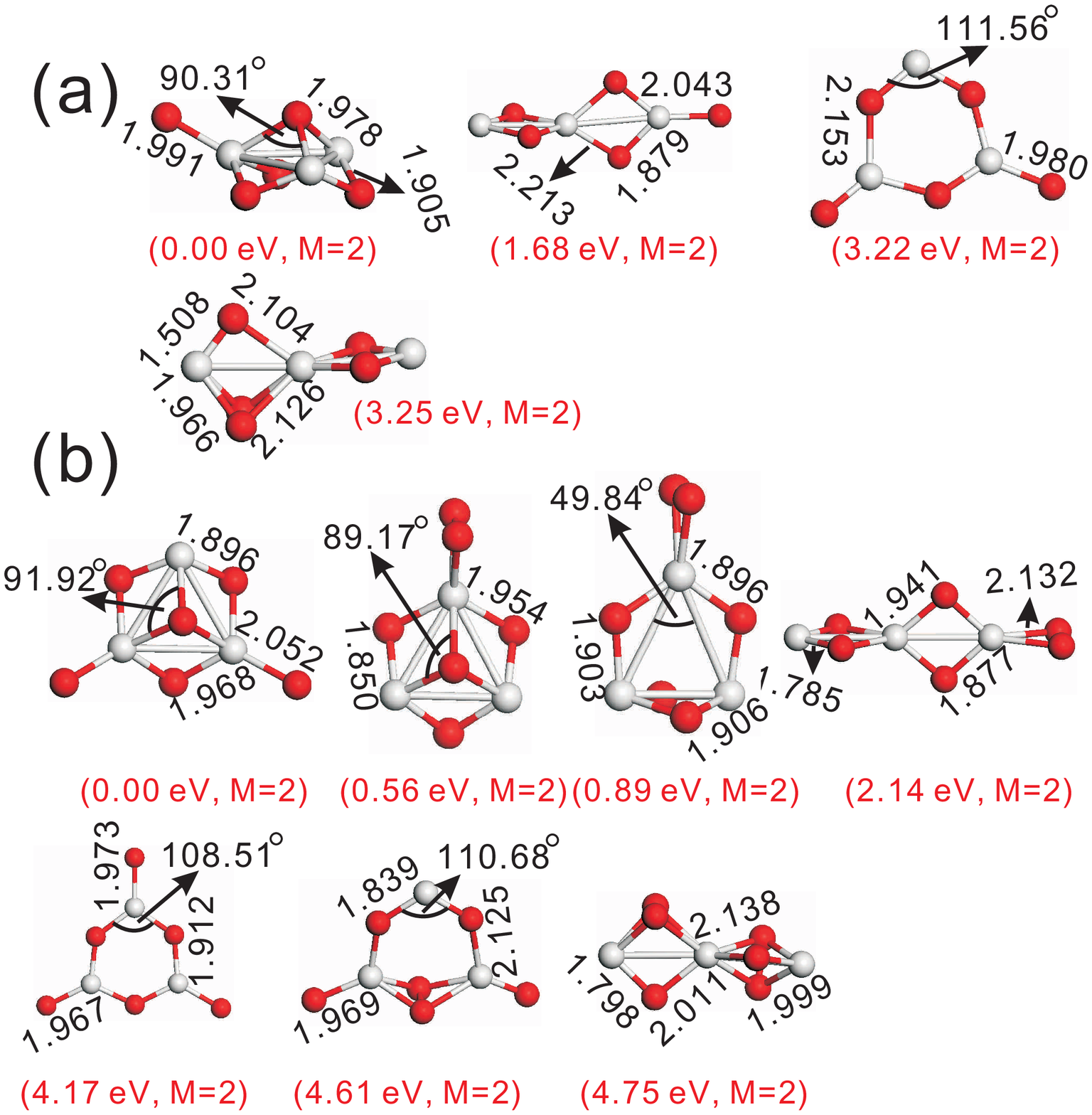}
\end{center}
\caption{(Color online). The low-energy structures of (a)
Sc$_3$O$_5$ and (b) Sc$_3$O$_6$ clusters. Grey and red balls
represent the scandium and oxygen atoms, respectively. The bond
lengths are in units of \AA. The number in parenthesis is the
relative energy (in eV) with respect to the corresponding ground
state. Note that the spin multiplicity is also given in the
parenthesis.}
\end{figure}%

As shown in Fig. 4(a), the ground-state structure for Sc$_3$O$_5$
can be seen as adding an oxygen atom on top of the scandium atom in
the ground-state structure of Sc$_3$O$_4$. The Sc-O bond lengths
become a little smaller than that in Sc$_3$O$_4$. The next stable
structure of Sc$_3$O$_5$ is the adsorption of 5 oxygen atoms on a
Sc$_3$ linear chain, whose free electronic energy is however, much
larger than the ground-state. The 3 lowest-energy structures of
Sc$_3$O$_6$ are the adsorption structures of 6 oxygen atoms on the
Sc$_3$ triangle. In the ground state, the oxygen atoms are separated
from each other, while in the next and third stable structure, two
of the six oxygen atoms bond together. We also find that the
structures contain a Sc$_3$ chain always have much larger free
electronic energies.

\subsection{Important features from the structural studies}

Based on the above systematic study of Sc$_n$O$_m$ clusters, we can
see some features for both the oxidation pattern of scandium, and
the influences of HSE calculation.

For the oxidation of scandium, we conclude two important features.
The first one is that the Sc$_n$O$_m$ cluster can be seen as adding
$m$ oxygen atoms to a Sc$_n$ cluster, with the Sc$_n$ cluster to be
scandium dimer for $n$=2, and scandium triangle for $n$=3. The other
feature is that the oxygen atoms in the ground-states of Sc$_n$O$_m$
clusters are all bondless with each other, only one O-O bond forms
in Sc$_2$O$_4$. The fact that oxygen atoms do not bond with each
other indicates that scandium oxide clusters are different from lead
oxide clusters, in which oxygen atoms can form O$_2$ or O$_3$ units
\cite{Liu07}. The structural features of Sc$_n$O$_m$ clusters tell
us that they can be built by adding $m$ oxygen atoms separately to a
Sc$_n$ cluster.

As for the improvements of the HSE method, the most important one is
that the symmetry of the cluster is often lower in HSE calculation
than in PBE calculation. The lengths of Sc-O bonds always differ
from each other in a cluster after geometry optimization using the
HSE method. The second feature concluded from the above geometrical
studies is that the HSE method does not change the relative
stability between different structures. It means that geometry
optimization using the PBE type functional will not yield wrong
ground-state structures for scandium oxide clusters. At last,
although the electronic structures in PBE and HSE calculations are
the same for most Sc$_n$O$_m$ clusters, we do see differences in the
ground-state electronic structures of Sc$_3$O$_2$ and Sc$_3$O$_3$.
In PBE calculations, the magnetic moment is 3 $\mu_B$ for
Sc$_3$O$_2$, and 1 $\mu_B$ for Sc$_3$O$_3$. And corresponding the
magnetic moments obtained from HSE calculations are 1 $\mu_B$ for
Sc$_3$O$_2$, and 3 $\mu_B$ for Sc$_3$O$_3$.

\subsection{Fragmentation channels and dissociation energies}

The stability of scandium oxide clusters with different sizes and
stoichiometries is required to illustrate the growth pattern of
various nanostructures and to understand even the oxidation behavior
of the pristine scandium clusters. On the other hand, the study of
the stability is helpful for finding the candidates of the building
block of the cluster-assembled materials. Therefore, in the
following, we evaluate the fragmentation energy of Sc$_n$O$_m$
clusters.

\begin{figure}[tbp]
\begin{center}
\includegraphics[width=0.8\linewidth]{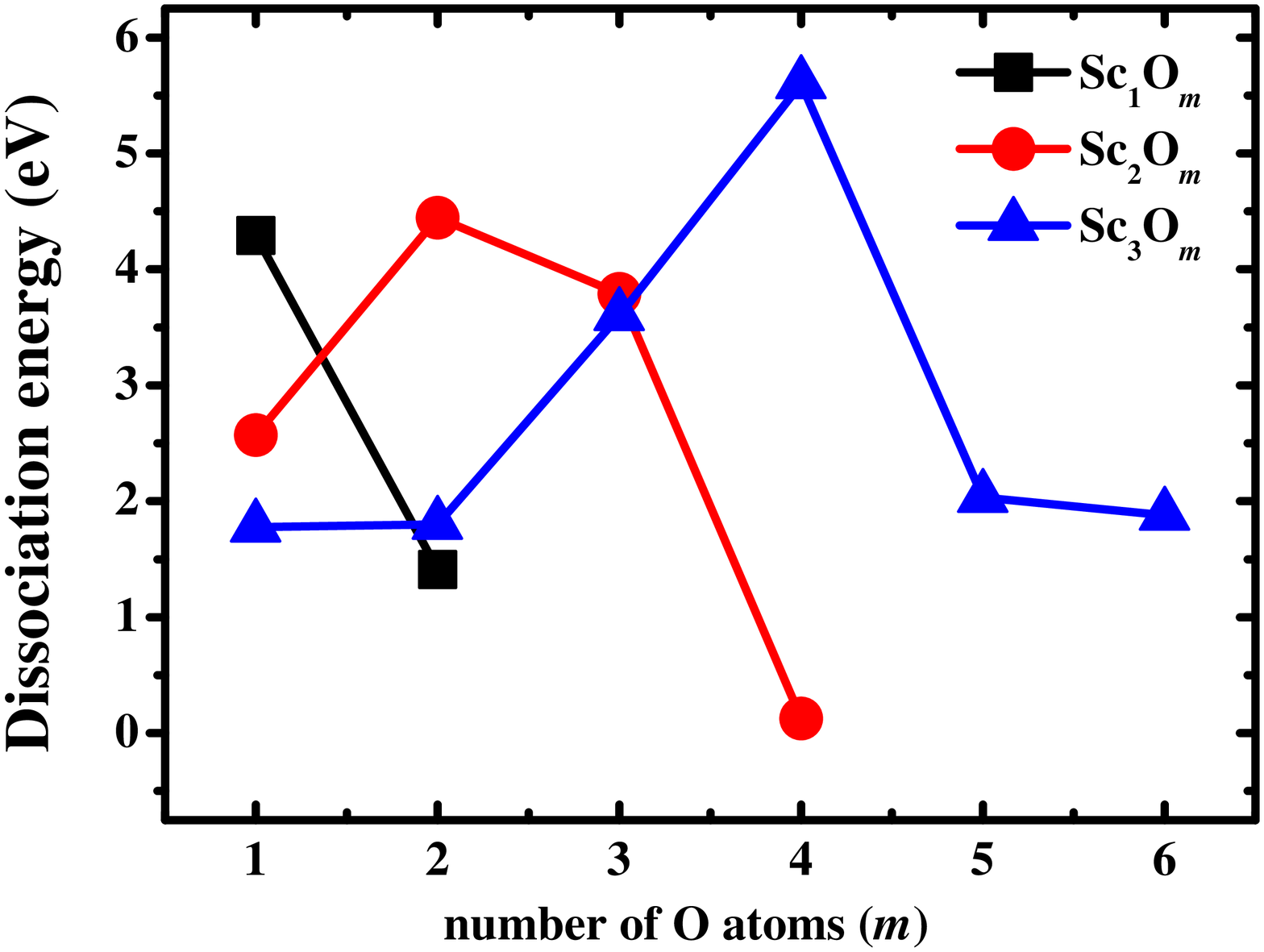}
\end{center}
\caption{(Color online). Dissociation energies of the Sc$_n$O$_m$
clusters as a function of the oxygen atom number $m$. The
corresponding fragmentation channels are listed in Table I.}
\end{figure}%

\begin{table}[ptb]
\caption{The most favorable fragmentation channels and dissociation
energies ($\Delta E$, in units of eV) of the Sc$_n$O$_m$ ($1\leq
n\leq 3,~1\leq m\leq 2n$) clusters. All results are obtained by
employing the HSE method.}
\begin{tabular}
[c] { l l l} \hline \hline
cluster ~~~~& Fragmentation channels ~~~~& $\Delta E$ \\
\hline
ScO         & Sc+O            & +4.29 \\
ScO$_2$     & ScO+O           & +1.41 \\
Sc$_2$O     & Sc+ScO          & +2.57 \\
Sc$_2$O$_2$ & ScO+ScO         & +4.45 \\
Sc$_2$O$_3$ & Sc$_2$O$_2$+O   & +3.79 \\
Sc$_2$O$_4$ & Sc$_2$O$_3$+O   & +0.12 \\
Sc$_3$O     & Sc+Sc$_2$O      & +1.78 \\
Sc$_3$O$_2$ & Sc+Sc$_2$O$_2$  & +1.80 \\
Sc$_3$O$_3$ & ScO+Sc$_2$O$_2$ & +3.60 \\
Sc$_3$O$_4$ & ScO+Sc$_2$O$_3$ & +5.60 \\
Sc$_3$O$_5$ & Sc$_3$O$_4$+O   & +2.03 \\
Sc$_3$O$_6$ & Sc$_3$O$_5$+O   & +1.88 \\
\hline \hline
\end{tabular}\label{Ead1}
\end{table}

When a cluster $A$ is dissociated into $B$ and $C$ fragments (i.e.,
$A\rightarrow B+C$), the fragmentation energy is defined as $\Delta
E =E_B+E_C-E_A$, where the $E_B$, $E_C$, and $E_A$ are the free
electronic energies of clusters $B$, $C$, and $A$, respectively. In
addition, the half of the total energy of an oxygen molecule is
adopted as the reference energy for atomic oxygen. Note that a
fragmentation process is exothermic (endothermic) if the associated
fragmentation energy is negative (positive). The dissociation energy
$\Delta E$, defined as the fragmentation energy of the most
favorable channel, is calculated and presented in Table I. In
general, the cluster with large positive $\Delta E$ has great
stability, and that with small positive or even negative $\Delta E$
is not stable and tends to dissociate. Moreover, the frequently
observed fragmentation products are believed to be stable
\cite{Liu07,Lu03}. Figure 5 shows the $\Delta E$ as a function of
the oxygen atom numbers for for all clusters, revealing the
underlying relationship between stability and stoichiometry of these
clusters. The essential features can be discussed as follows.

Firstly, for the dissociation of Sc-rich clusters, the Sc atom is
always one of the fragments. To make a further validation, we also
study possible fragmentation ways for the Sc$_4$O cluster, using the
HSE calculational method. It is found that the most energetically
favored fragmentation channel is Sc$_4$O$\rightarrow$Sc+Sc$_3$O, in
which single Sc atom is also a fragment. Secondly, we do not see
oxygen molecules in the fragments of Sc$_n$O$_m$ clusters, the
O-rich clusters favor the fragmentation channels that have a single
oxygen atom as a product. Lastly, the dissociation energies of ScO,
Sc$_2$O$_2$, Sc$_2$O$_3$, Sc$_3$O$_3$, and Sc$_3$O$_4$ clusters are
larger than 3.60 eV. This indicates that the monoxide-like and
sesquioxide-like scandium oxide clusters, and oxide clusters between
them are remarkably stable. As a result, these clusters are
frequently observed in the fragmentation channels shown in Table I.
We also notice that the Sc$_3$O$_4$ cluster has an enormous
dissociation energy, and is more stable than the monoxide-like
Sc$_3$O$_3$ cluster. And at another side, the monoxide-like
Sc$_2$O$_2$ cluster, with a larger dissociation energy, is more
stable than the the sesquioxide-like Sc$_2$O$_3$ cluster.

\subsection{Electronic energy levels}

The electronic structures of transition metal oxide clusters are
important characters for their experimental detections and chemical
applications. Especially, their magnetic properties are very
important to reflect the behaviors of the $d$-shell electrons. We
thus calculate the ground-state electronic energy levels of
Sc$_n$O$_m$ clusters, using the HSE method. Figure 6 shows the
obtained energy levels for different spin polarized electrons, in
which spin-up and spin-down represent for the majority and minority
spin states respectively. We find that the Sc$_n$O$_m$ clusters with
odd-number Sc atoms are all magnetic. For the Sc$_2$O$_m$ clusters,
while the Sc$_2$O is magnetic, the other Sc$_2$O$_2$, Sc$_2$O$_3$,
and Sc$_2$O$_4$ clusters are nonmagnetic. The energy gap between the
highest occupied molecular orbital (HOMO) and the lowest unoccupied
molecular orbital (LUMO) is 1.82, 3.36, and 3.20 eV for the
Sc$_2$O$_2$, Sc$_2$O$_3$, and Sc$_2$O$_4$ cluster respectively. The
large band gaps indicate that the Sc$_2$O$_2$, Sc$_2$O$_3$ and
Sc$_2$O$_4$ clusters are chemically very stable. In contrast,
although the Sc$_3$O$_4$ cluster is energetically very stable, its
energy gap of spin-up electrons is as small as 0.49 eV, indicating
that it is chemically active, and ready to interact with other
particles or materials.

\begin{figure}[tbp]
\begin{center}
\includegraphics[width=0.8\linewidth]{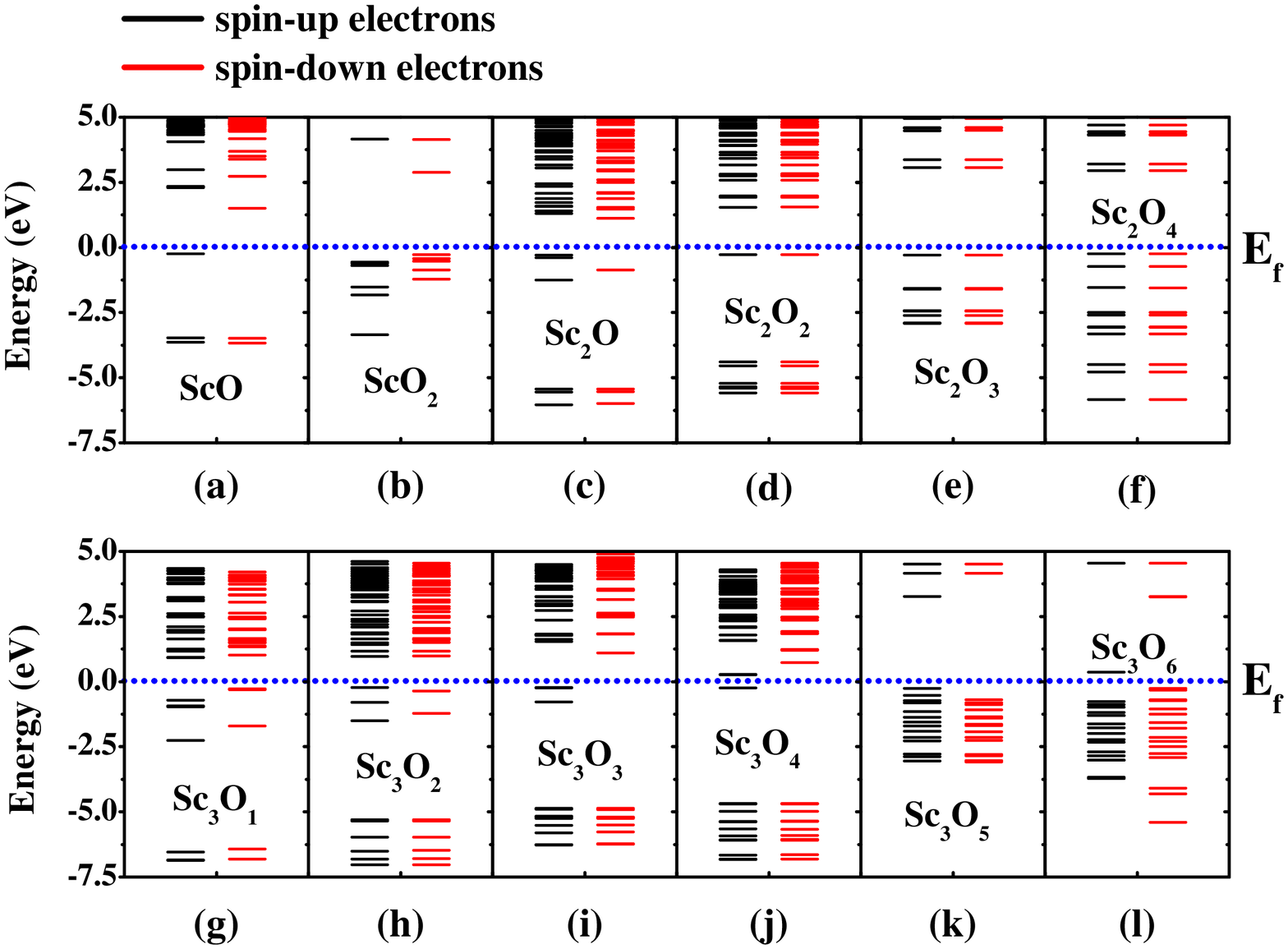}
\end{center}
\caption{(Color online). The electronic energy levels for (a) ScO,
(b) ScO$_2$, (c) Sc$_2$O, (d) Sc$_2$O$_2$, (e) Sc$_2$O$_3$, (f)
Sc$_2$O$_4$, (g) Sc$_3$O, (h) Sc$_3$O$_2$, (i) Sc$_3$O$_3$, (j)
Sc$_3$O$_4$, (k) Sc$_3$O$_5$, and (l) Sc$_3$O$_6$ clusters, obtained
from HSE calculations. Spin-up and spin-down electronic states are
denoted by black and red lines, respectively. The Fermi levels are
denoted by the blue dotted lines.}
\end{figure}%

\begin{figure}[tbp]
\begin{center}
\includegraphics[width=0.3\textwidth]{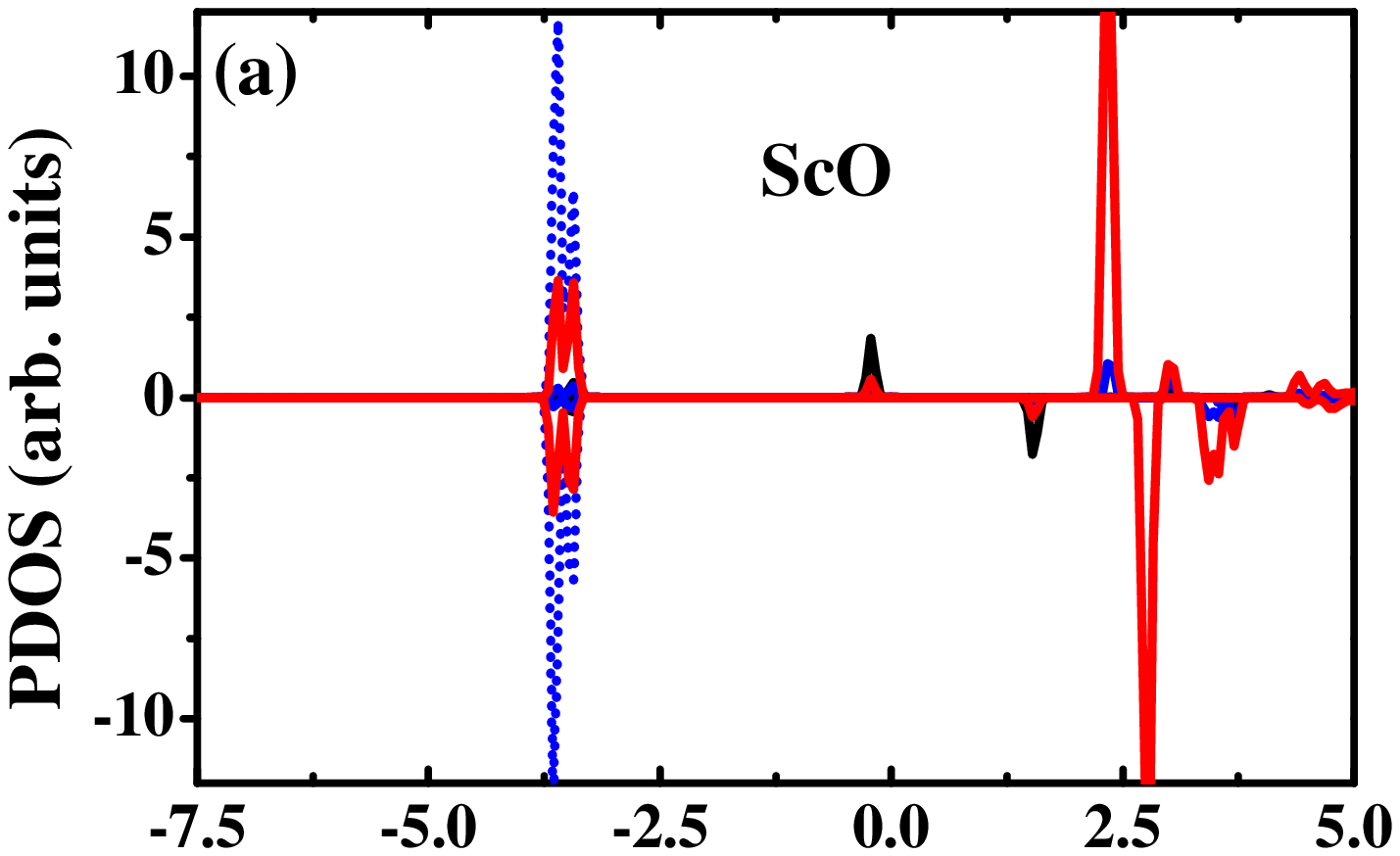}
\includegraphics[width=0.3\textwidth]{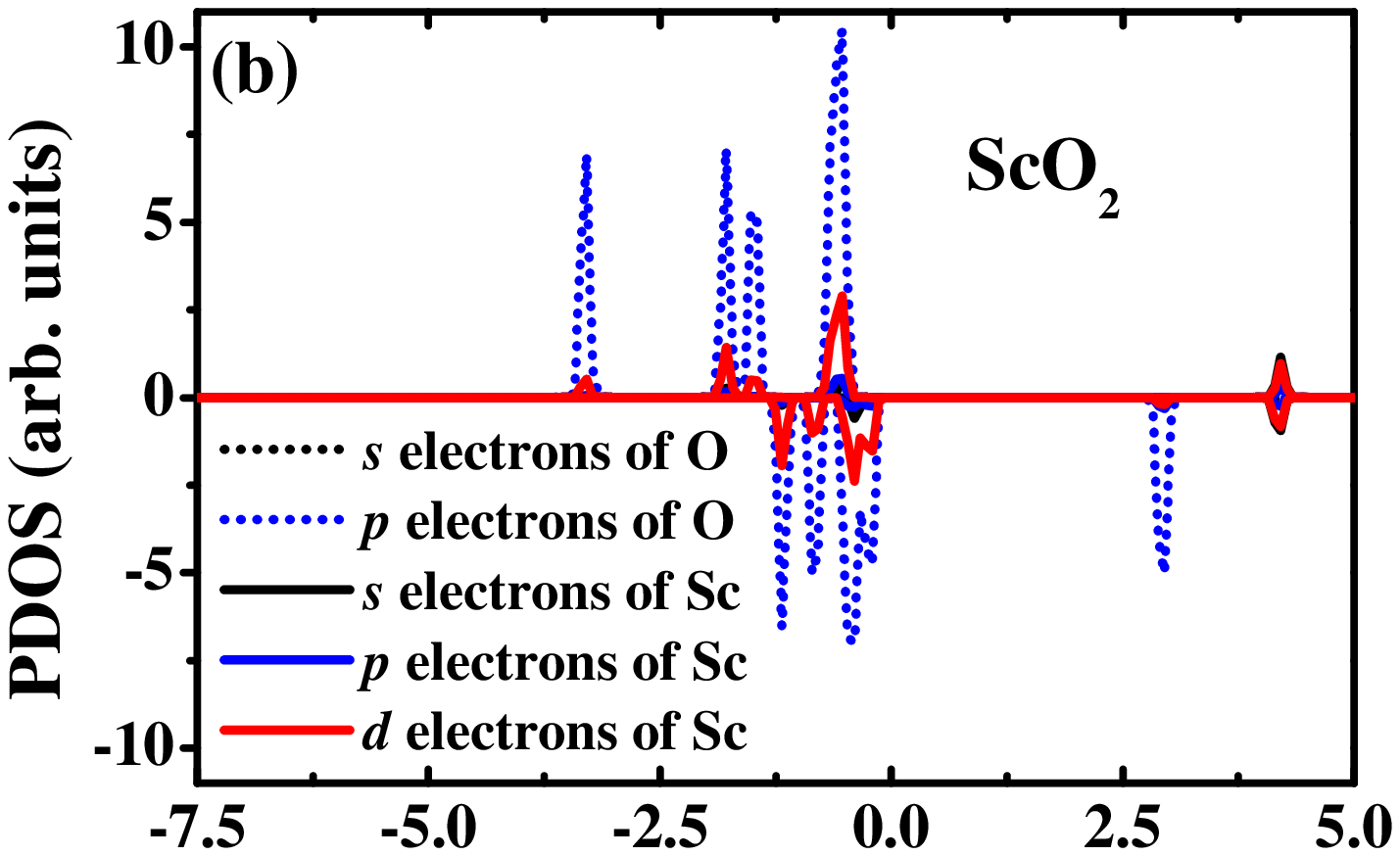}
\includegraphics[width=0.3\textwidth]{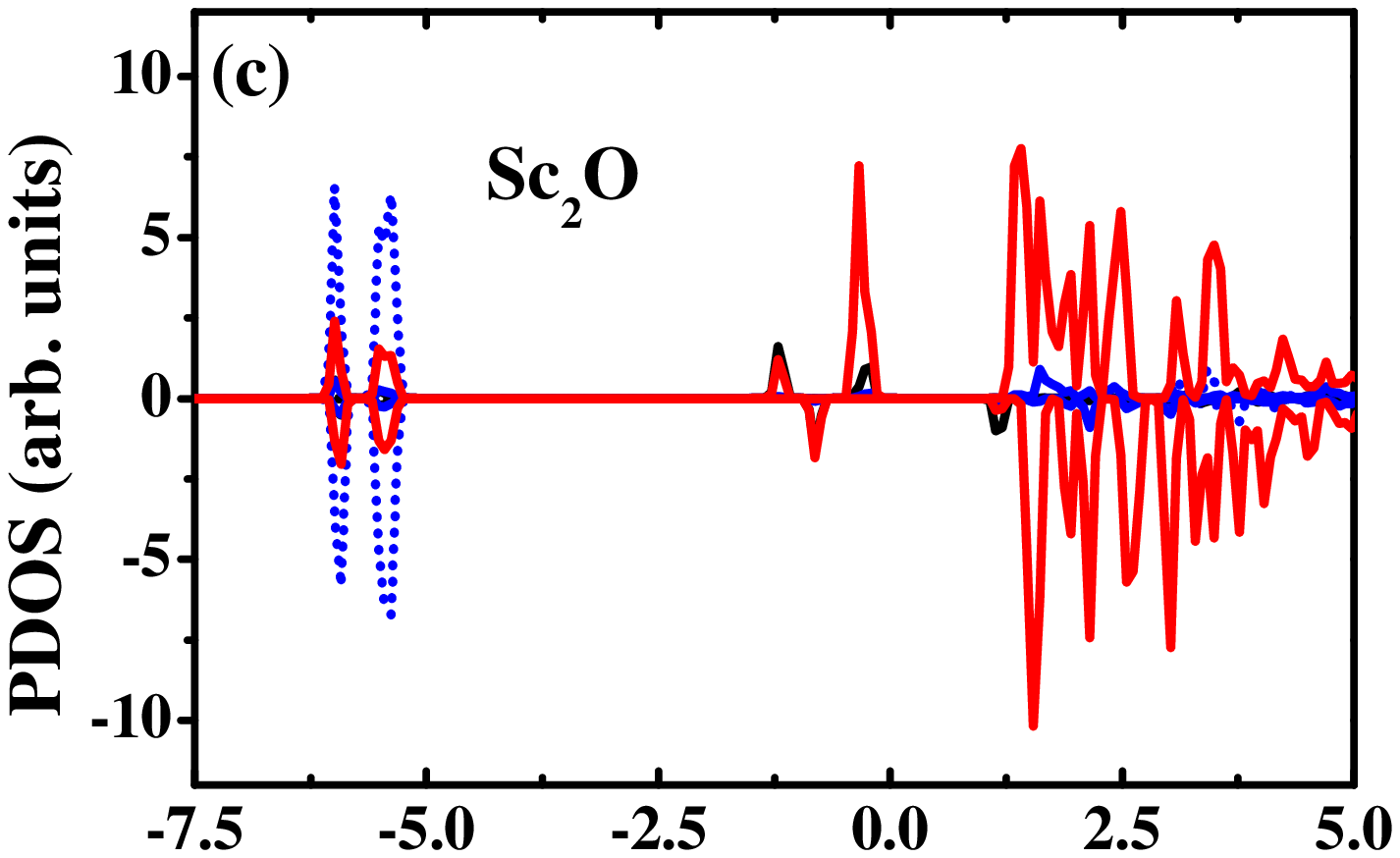}
\includegraphics[width=0.3\textwidth]{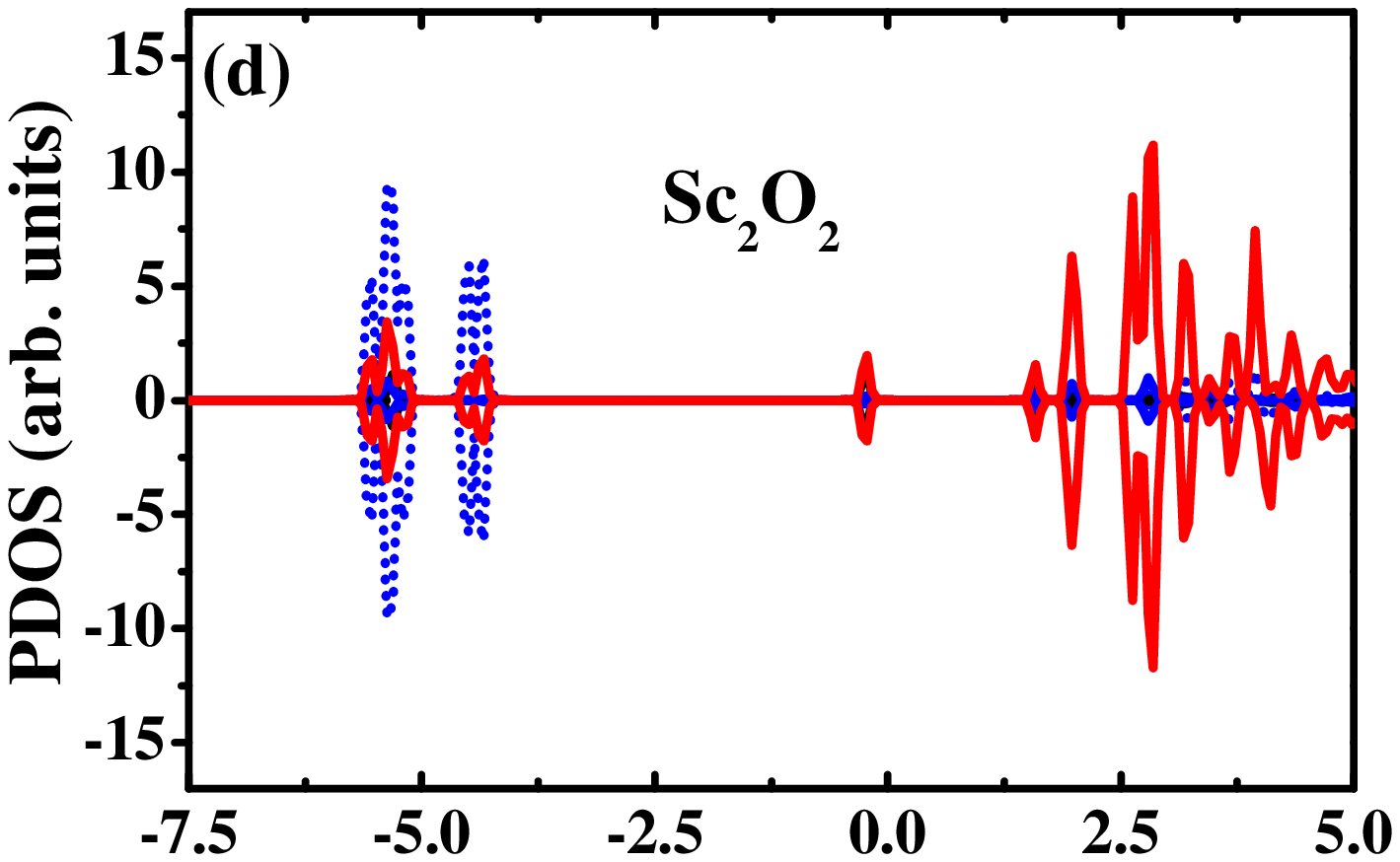}
\includegraphics[width=0.3\textwidth]{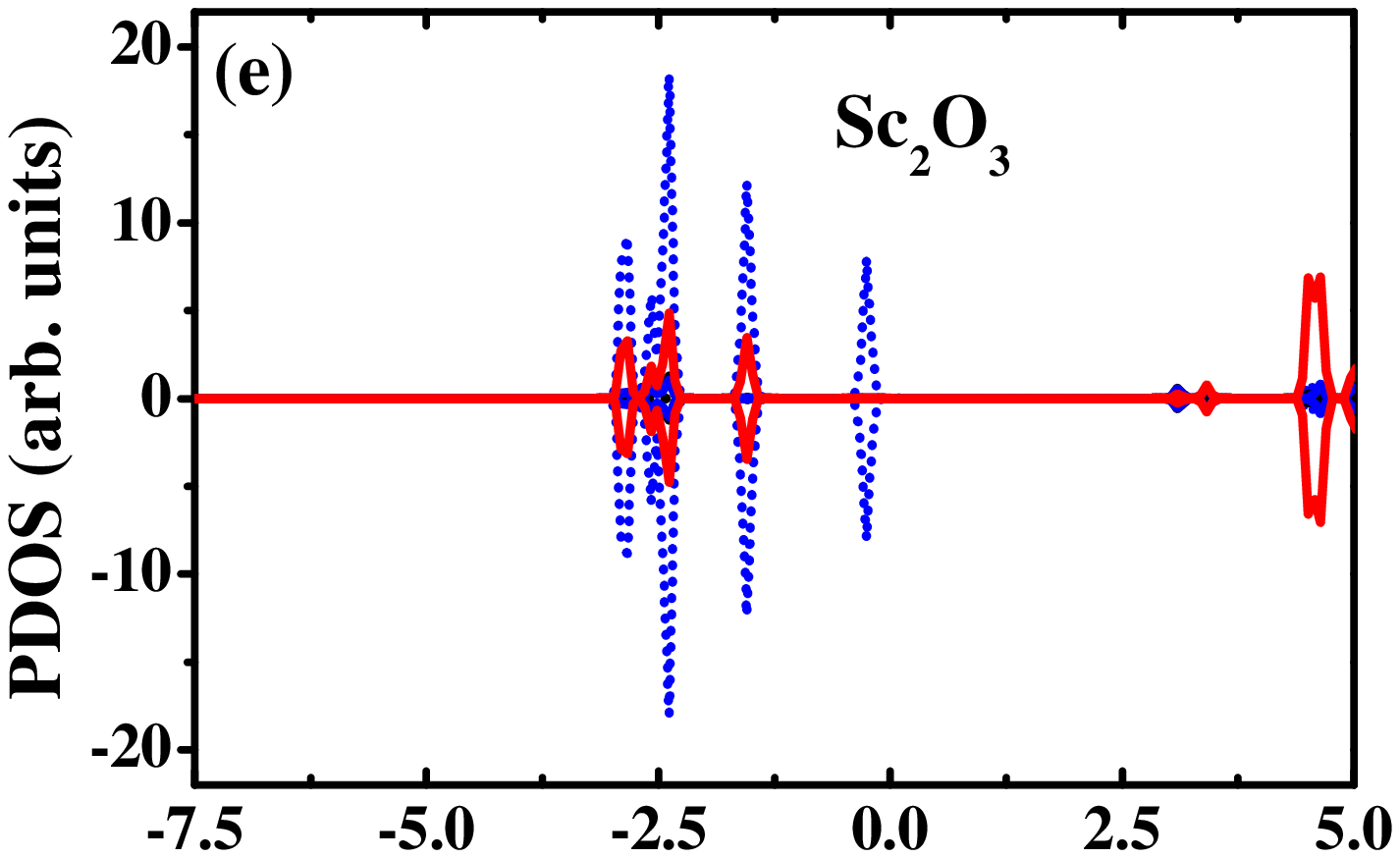}
\includegraphics[width=0.3\textwidth]{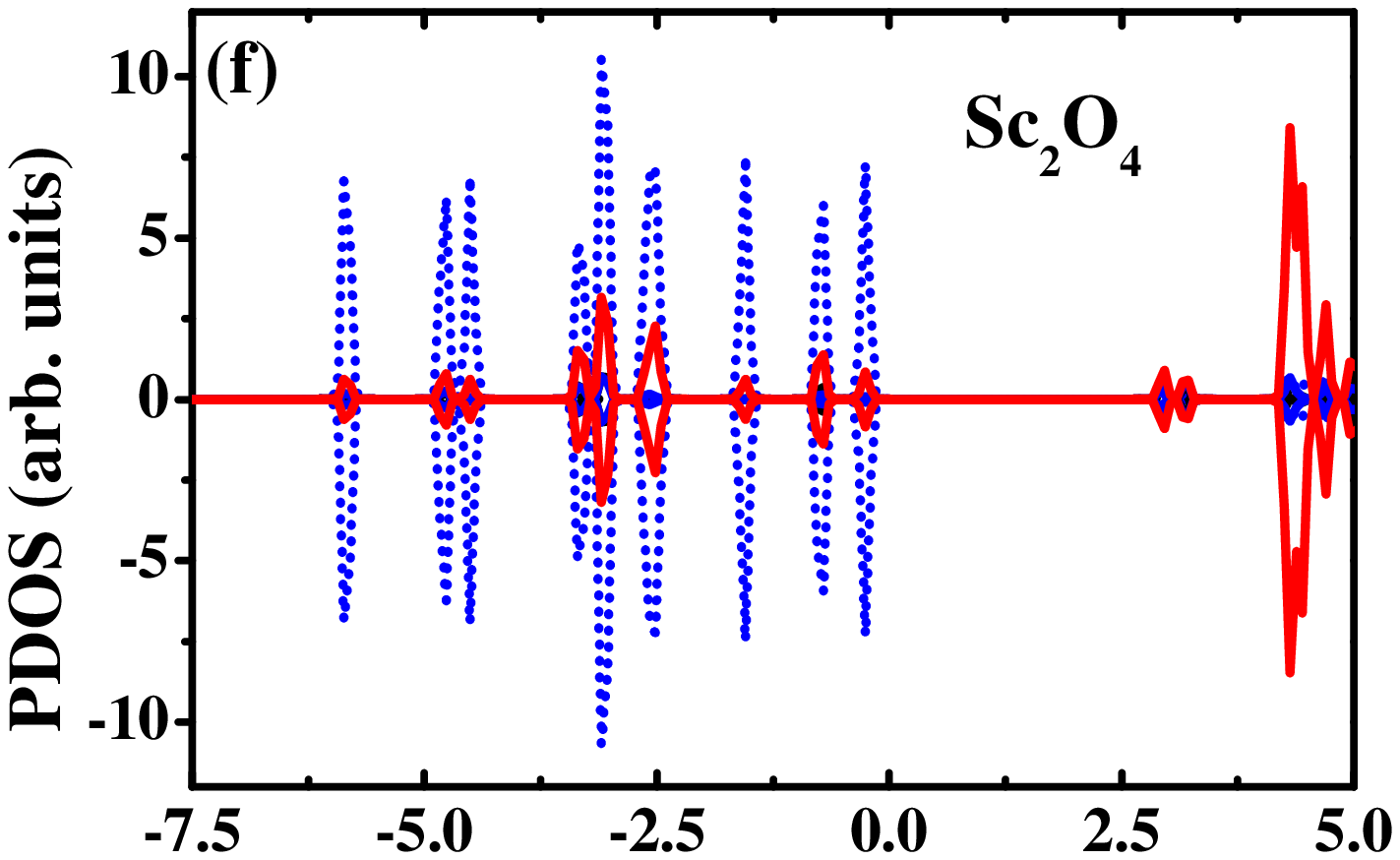}
\includegraphics[width=0.3\textwidth]{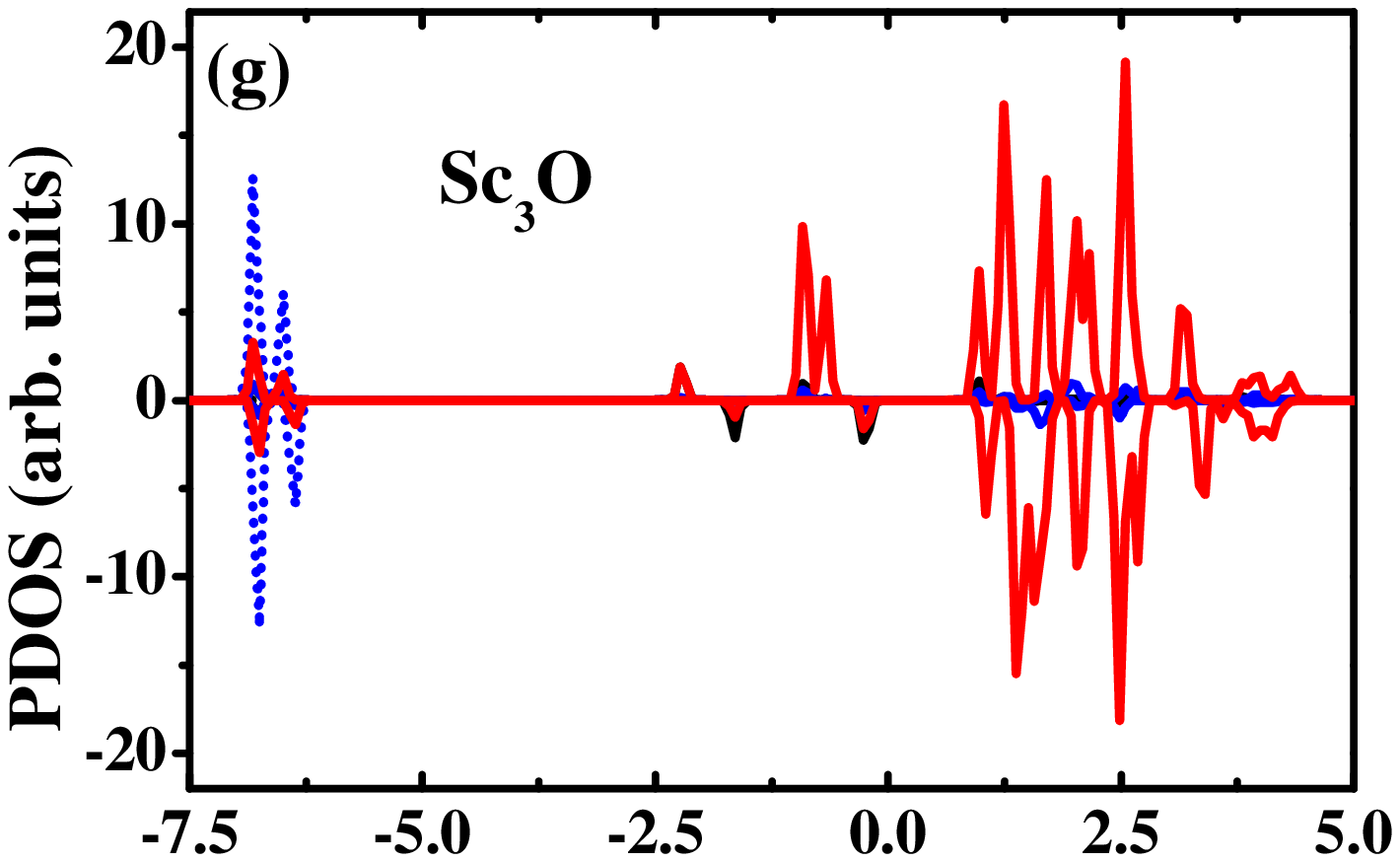}
\includegraphics[width=0.3\textwidth]{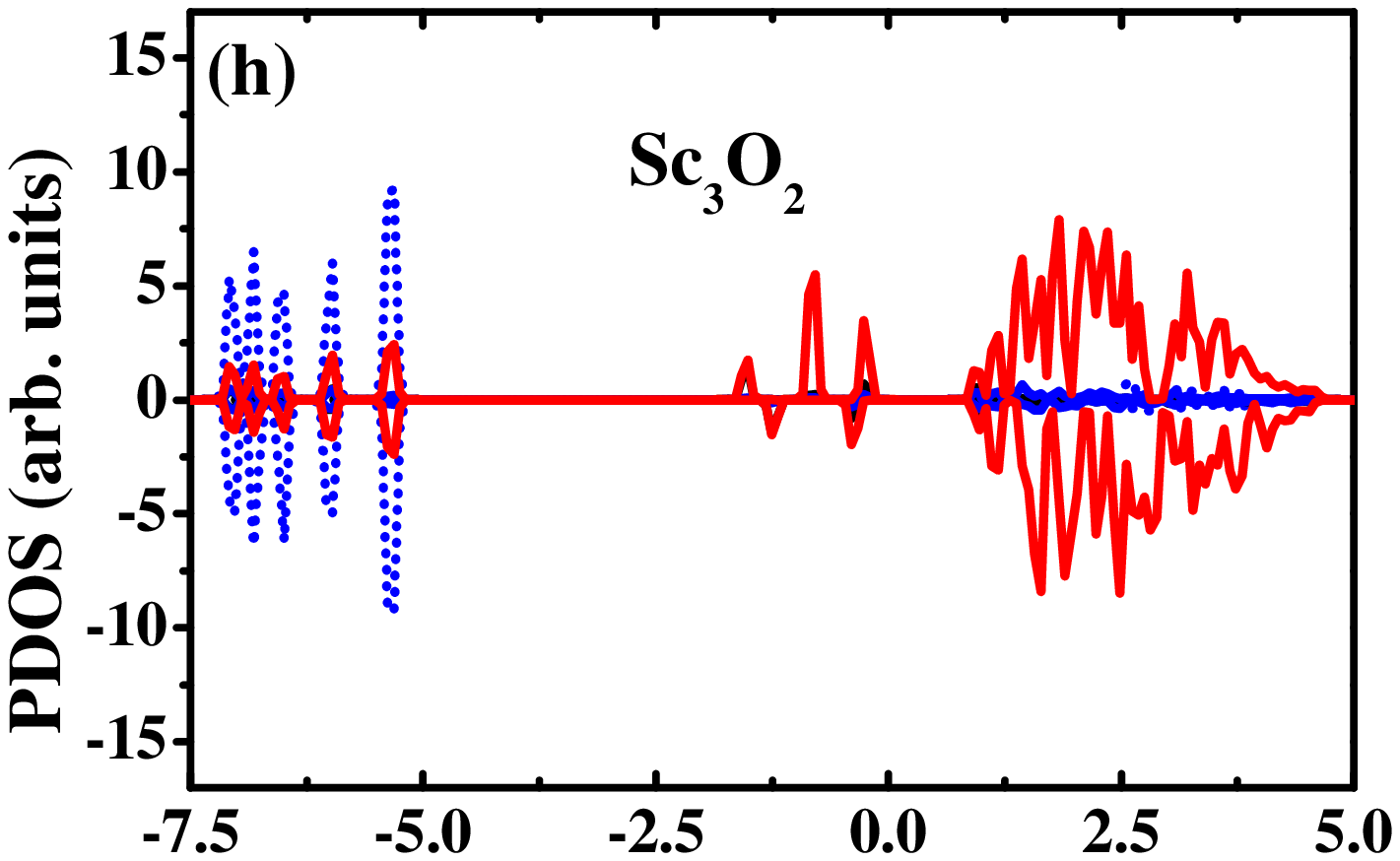}
\includegraphics[width=0.3\textwidth]{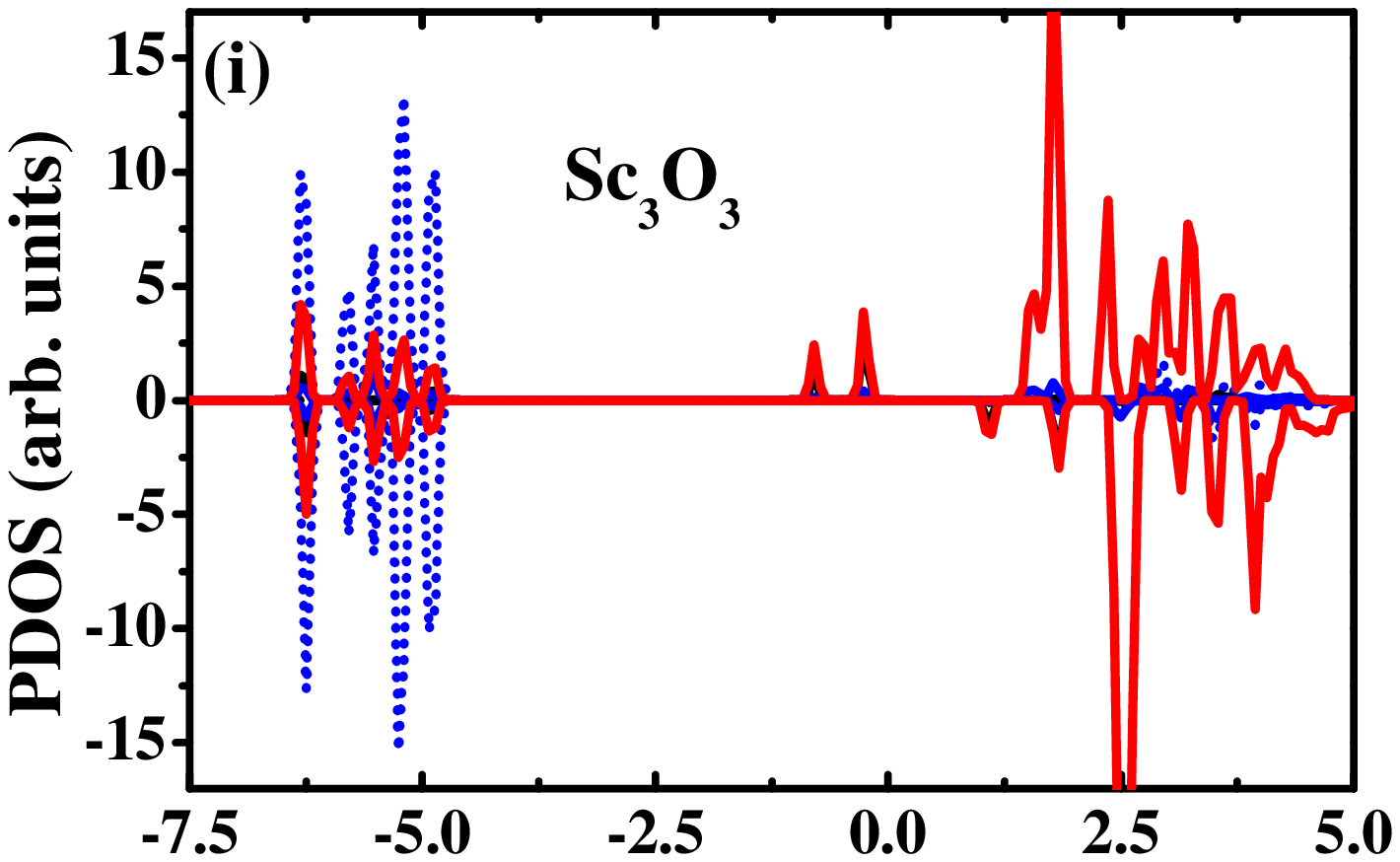}
\includegraphics[width=0.3\textwidth]{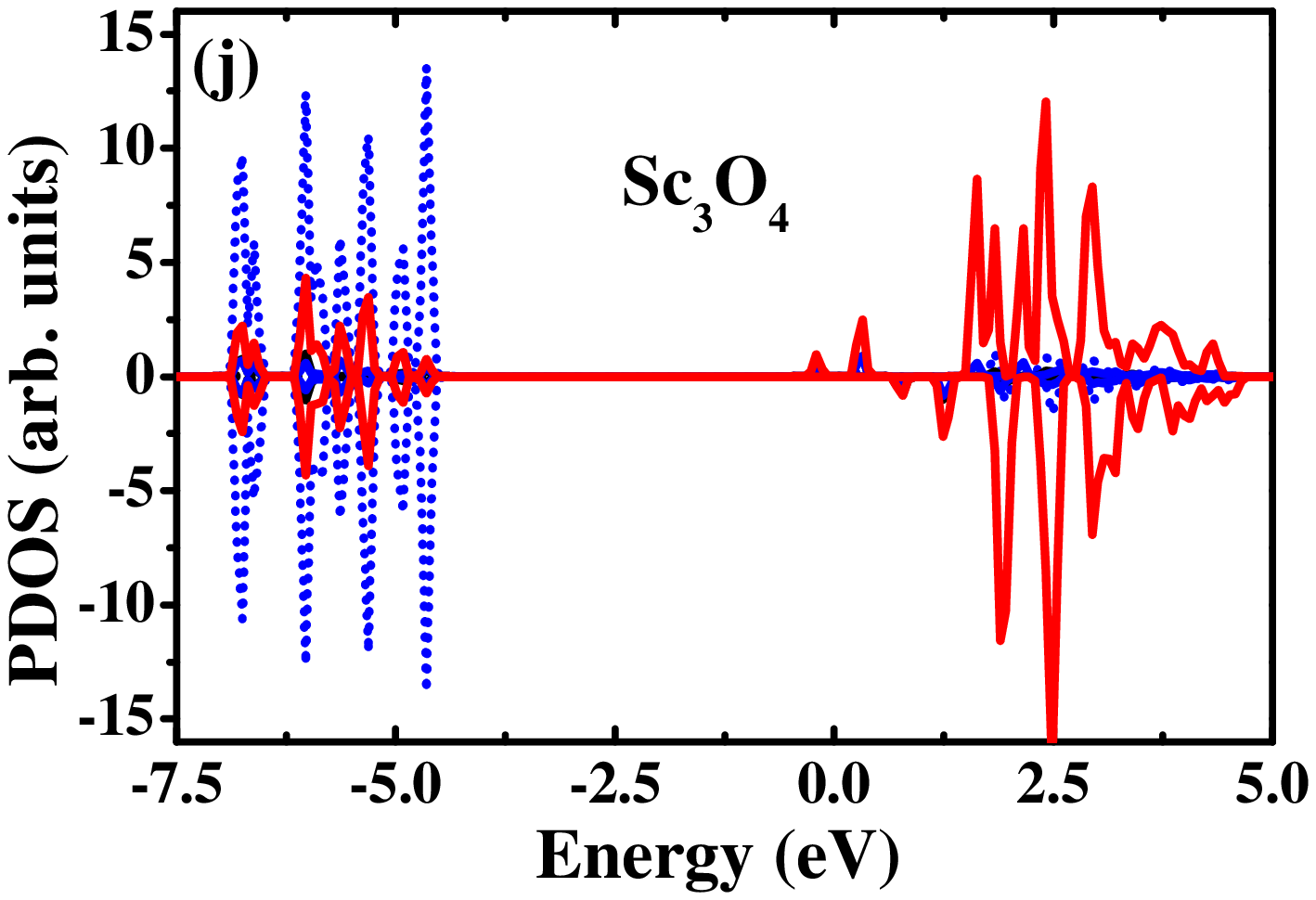}
\includegraphics[width=0.3\textwidth]{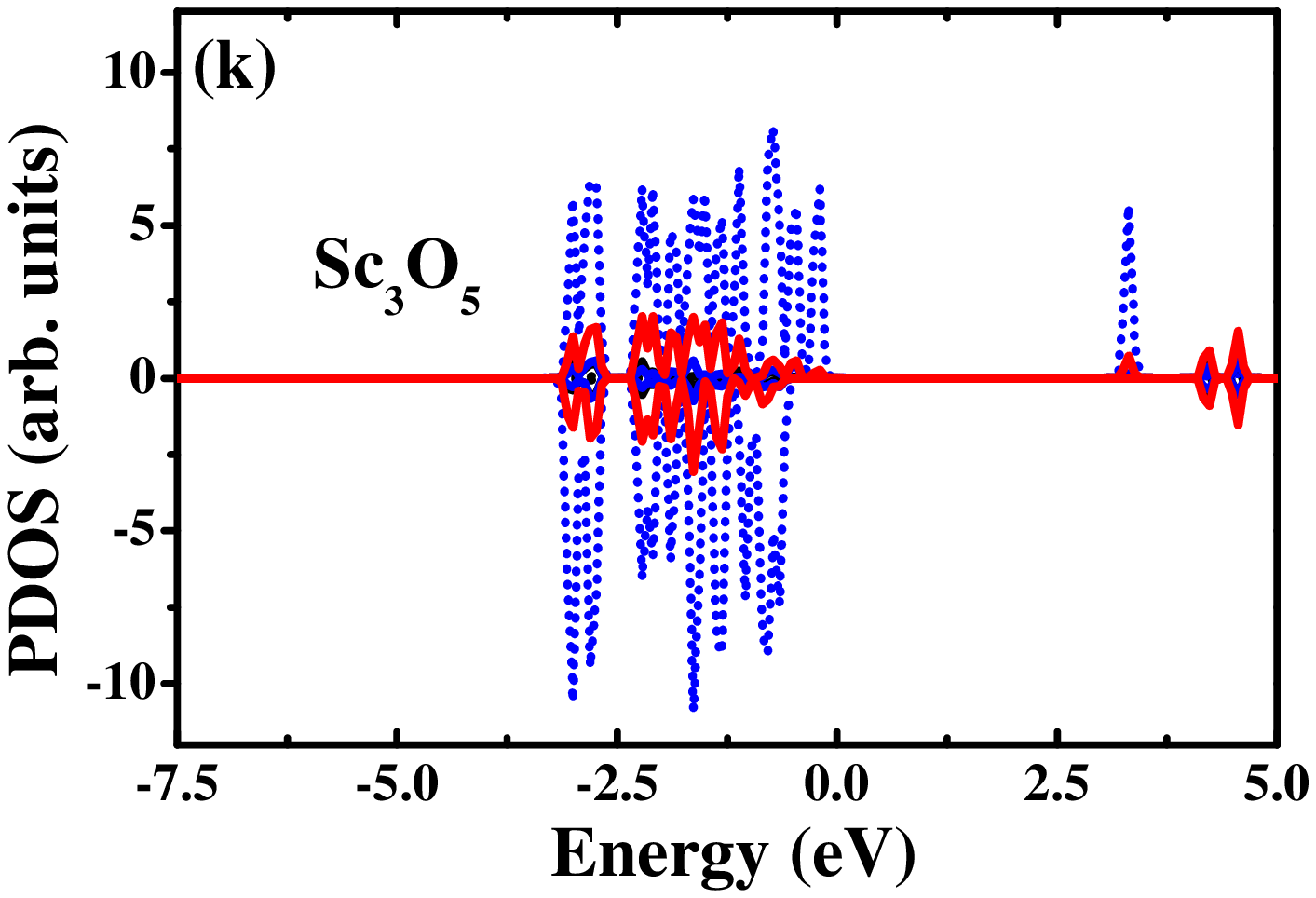}
\includegraphics[width=0.3\textwidth]{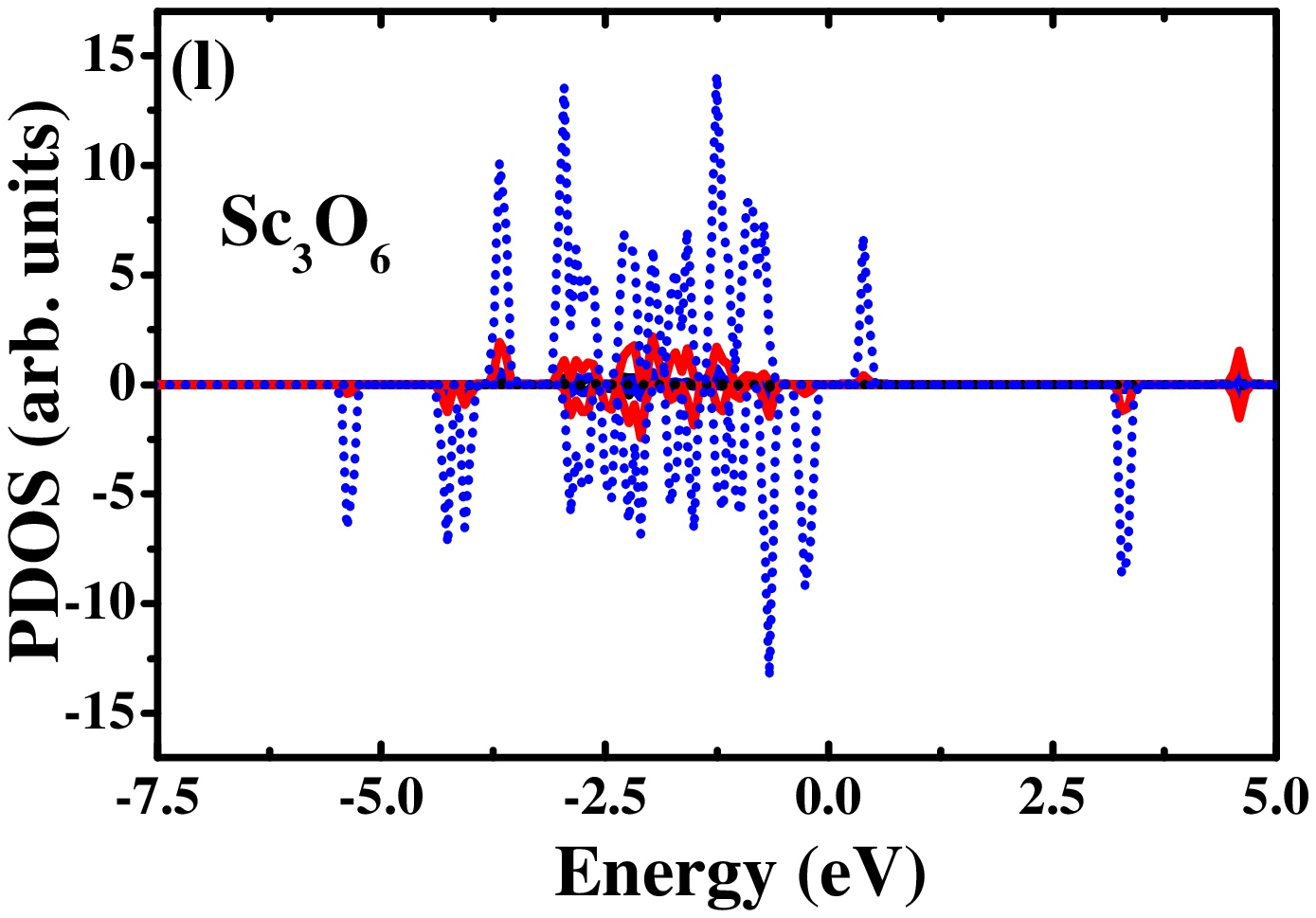}
\end{center}
\caption{(Color online). The spin-resolved projected density of
states for the (a) ScO, (b) ScO$_2$, (c) Sc$_2$O, (d) Sc$_2$O$_2$,
(e) Sc$_2$O$_3$, (f) Sc$_2$O$_4$, (g) Sc$_3$O, (h) Sc$_3$O$_2$, (i)
Sc$_3$O$_3$, (j) Sc$_3$O$_4$, (k) Sc$_3$O$_5$, and (l) Sc$_3$O$_6$
clusters. The Fermi energies are all set to be zero. The electronic
states of O and Sc are shown in dotted and solid line respectively,
while $s$, $p$ and $d$ electronic states are shown as black, blue
and red lines respectively.}
\end{figure}%

To systematically investigate the electronic structures of
Sc$_n$O$_m$ clusters, we then calculate the projected density of
states (PDOS) for the Sc$_n$O$_m$ clusters, which are spin-resolved.
Figures 7(a)-7(l) list the obtained $s$-, $p$-, and $d$-PDOS of
oxygen and scandium atoms, for the Sc$_n$O$_m$ clusters
respectively. We can clearly see that for all the studied
Sc$_n$O$_m$ clusters, there are strong hybridizations between oxygen
2$p$ and scandium 3$d$ electronic states, and they contribute most
of the electronic states around the Fermi energies. For Sc-rich
clusters like Sc$_2$O, Sc$_3$O, and  Sc$_3$O$_2$, the HOMO and LUMO
are both composed of Sc-3$d$ electrons. For the three nonmagnetic
Sc$_2$O$_2$, Sc$_2$O$_3$, and Sc$_2$O$_4$ clusters, the HOMO and
LUMO are contributed by O-2$p$ electrons and Sc-3$d$ electrons,
respectively. And for O-rich clusters like ScO$_2$, Sc$_3$O$_5$, and
Sc$_3$O$_6$, the HOMO and LUMO are both 2$p$ electronic states of
oxygen. The electronic structure of ScO is, however, different from
the other clusters, for its HOMO of spin-up electrons and LUMO of
spin-down electrons are contributed by 4$s$ electronic states of
scandium.

\begin{figure}[tbp]
\begin{center}
\includegraphics[width=0.8\linewidth]{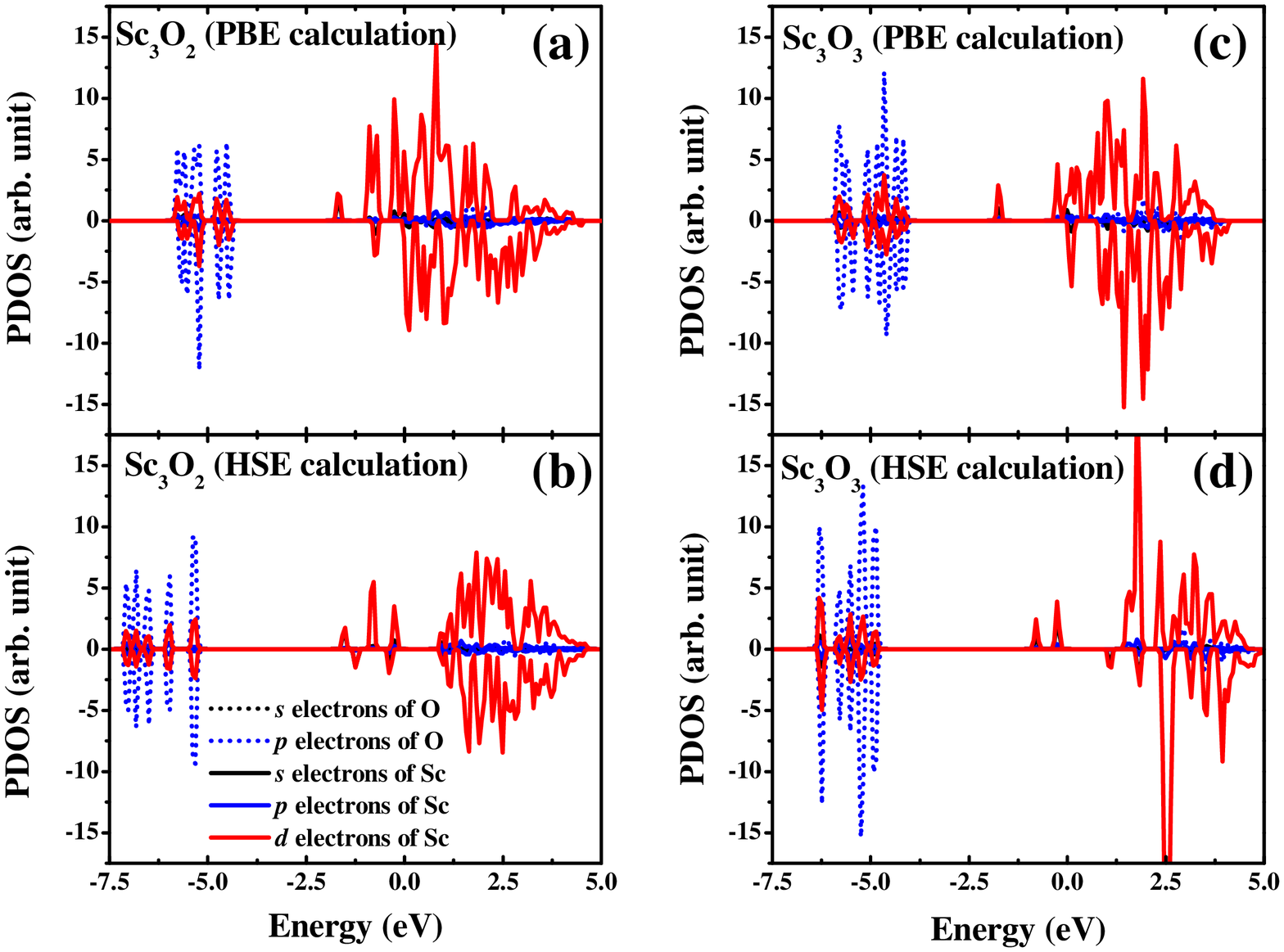}
\end{center}
\caption{(Color online). The spin-resolved projected density of
states for the Sc$_3$O$_2$ cluster in (a) PBE and (b) HSE
calculations, and the Sc$_3$O$_3$ cluster in (c) PBE and (d) HSE
calculations. The Fermi energies are all set to be zero. The
electronic states of O and Sc are shown in dotted and solid line
respectively, while $s$, $p$ and $d$ electronic states are shown as
black, blue and red lines respectively.}
\end{figure}%

Since the obtained ground-state spin configurations are different
for the Sc$_3$O$_2$ and Sc$_3$O$_3$ clusters in PBE and HSE
calculations, we draw their PDOS together to compare the differences
in electronic-state descriptions of PBE and HSE methods, which are
shown in Figs. 8(a)-8(d). In the $p$-$d$ hybridization area from
-7.0 eV to -3.0 eV below the Fermi energies, we can see that the
hybridized peaks are more separate in HSE calculations for both the
Sc$_3$O$_2$ and Sc$_3$O$_3$ clusters. The electronic states around
the Fermi energy are both Sc-3$d$ states for the two chosen
clusters. One can see from Fig. 8 that there are always more
localized Sc-3$d$ peaks around the Fermi energy in HSE calculations
than in PBE calculations. It means that the HSE method can lead to
more localized descriptions for Sc-3$d$ electronic states in
Sc$_n$O$_m$ clusters. Considering that the PBE and other standard
GGA methods rely on the $xc$ energy of the uniform electron gas, and
thus are expected to be useful only for systems with slowly varying
electron densities, we think that the HSE descriptions on the
Sc-3$d$ states are more reasonable.

\section{Conclusions}

In this work, first-principles calculations with the HSE method have
been performed to study the geometries, stabilities and electronic
structures of small Sc$_n$O$_m$ clusters ($n$=1-3,$m$=1-2$n$). Based
on an extensive search, it is found that the lowest-energy
structures of all these clusters can be obtained by the sequential
oxidation of small "core" scandium clusters, with the adsorbing
oxygen atoms separating from each other.

The fragmentation analysis reveals that the ScO, Sc$_2$O$_2$,
Sc$_2$O$_3$, Sc$_3$O$_3$, and Sc$_3$O$_4$ clusters have great
stability. This suggests that these clusters might be used as
candidates of the building block of cluster-assembled materials. The
fragmentation of Sc-rich clusters is found to include a scandium
atom as a product, while the fragmentation of O-rich clusters is
found to include an oxygen atom as a product. Besides, the above
four extremely stable clusters can also be frequently seen in the
fragmentation products of other Sc$_n$O$_m$ clusters.

Through electronic structure calculations and wavefunction analysis,
we reveal that most Sc oxide clusters have magnetic ground state
except Sc$_2$O$_2$, Sc$_2$O$_3$, and Sc$_2$O$_4$ clusters. The HOMO
and LUMO of the three nonmagnetic clusters are composed of O-2$p$
and Sc-3$d$ electronic states, respectively. For the Sc-rich
clusters, the HOMO and LUMO are contributed both by Sc-3$d$
electrons, while for O-rich clusters, the HOMO and LUMO are
contributed by O-2$p$ electrons.

In comparison with standard PBE calculations, the HSE method is
superior because it can correct the wrong symmetries and electronic
configurations in PBE results of some clusters.

\begin{acknowledgments}
This work was supported by the NSFC under Grants No. 90921003 and
No. 10904004.
\end{acknowledgments}

\end{document}